\documentclass[epj-spec]{svjour}
\usepackage{graphics}
\usepackage{epsfig}
\usepackage{amsmath}
\usepackage{amssymb}
\usepackage{epsf}
\usepackage{graphicx}
\usepackage{cite}
\usepackage{rotating}
\usepackage{slashed}
\usepackage{upgreek}
\usepackage{xcolor}
\usepackage{hyperref}
\hypersetup{colorlinks=true,linkbordercolor=red,linkcolor=blue, citecolor=blue,pdfborderstyle={/S/U/W 1}}
\usepackage{url}

\def\url#1{\href{#1}{{\tt #1}}}

\DeclareMathSymbol{\Upsilon}{\mathalpha}{operators}{7}
\DeclareMathSymbol{\psi}{\mathalpha}{letters}{32}
\let\psi\uppsi

\newcommand{\ycrit}{y^{\rm crit}}

\newcommand{\eq}[1]{(\ref{#1})}

\newcommand{\Ep}{E_\mathrm{p}}
\newcommand{\mprot}{m_\mathrm{p}}
\newcommand{\mT}{M_\perp}
\newcommand{\jpsi}{{\mathrm J}/\psi}
\newcommand{\xf}{x_{\mathrm{F}}}
\newcommand{\pt}{p_{_\perp}}
\newcommand{\dd}{{\rm d}}

\newcommand{\qzero}{\hat{q}_0}

\newcommand{\gevsqfm}{GeV$^2$/fm}

\newcommand{\be}{\begin{equation}}
\newcommand{\ee}{\end{equation}}
\newcommand{\bea}{\begin{eqnarray}}
\newcommand{\eea}{\end{eqnarray}}

\newcommand{\ellb}{L}
\newcommand{\qhat}{\hat{q}}

\newcommand{\rpa}{R_{\rm pA}}

\newcommand{\dyb}{{\delta y}}
\newcommand{\dymax}{{\delta y^{\rm max}}}
\newcommand{\Ea}{E}
\newcommand{\Phat}{\hat{{\cal P}}}

\newcommand{\epsa}{\varepsilon}
\newcommand{\epsamax}{\varepsilon^{\rm max}}

\newcommand{\sqrts}{\sqrt{s}}
\newcommand{\ylab}{y_{_{\rm lab}}}
\newcommand{\Mt}{M_{\perp}}
\newcommand{\thard}{t_{\rm h}}

\def\pt{p_{_\perp}}

\begin{document}
\title{Quarkonium suppression from coherent energy loss\\[0.25cm] in fixed-target experiments using LHC beams}

\author{Fran\c{c}ois Arleo$^{1,}$\footnote{On leave from Laboratoire d'Annecy-le-Vieux de Physique Th\'eorique (LAPTh), UMR5108, Universit\'e de Savoie, CNRS, BP 110, 74941 Annecy-le-Vieux cedex, France}
        \and
        St\'ephane Peign\'e$^{2}$ 
}

\institute{Laboratoire Leprince-Ringuet (LLR), \'Ecole polytechnique, CNRS/IN2P3 91128 Palaiseau, France \label{addr1}
           \and
           SUBATECH, UMR 6457, Universit\'e de Nantes, Ecole des Mines de Nantes, IN2P3/CNRS \\ 4 rue Alfred Kastler, 44307 Nantes cedex 3, France \label{addr2}
}

\date{Received: date / Revised version: date}
\abstract{Quarkonium production in proton-nucleus collisions is a powerful tool to disentangle cold nuclear matter effects. A model based on coherent energy loss is able to explain the available quarkonium suppression data in a broad range of rapidities, from fixed-target to collider energies, suggesting cold energy loss to be the dominant effect in quarkonium suppression in p--A collisions. This could be further tested in a high-energy fixed-target experiment using a proton or nucleus beam. The nuclear modification factors of $\jpsi$ and $\Upsilon$ as a function of rapidity are computed in p--A collisions at $\sqrts=114.6$~GeV, and in p--Pb and Pb--Pb collisions at $\sqrts=72$~GeV. These center-of-mass energies correspond to the collision on fixed-target nuclei of 7~TeV protons and 2.76~TeV~lead nuclei available at the LHC.}
\maketitle

\section{Introduction}

Understanding the physical origin of quarkonium ($\jpsi$, $\Upsilon$) suppression in proton--nucleus (p--A) collisions has been a challenge for the past thirty years. This would of course be a prerequisite in order to perform reliable \emph{baseline} predictions in heavy-ion collisions, where quarkonia are expected to be dissociated due to Debye screening of the heavy-quark potential at finite temperature~\cite{Matsui:1986dk}. Perhaps even more importantly, the wealth of data 
(especially for $\jpsi$ and $\Upsilon$) 
available in p--A collisions could help to understand 
generic features of 
hard QCD processes in a nuclear environment.

In everyday language, we often make the distinction between `fixed-target' and `collider' experiments when it comes to quarkonium production. 
This separation might look a bit artificial but not entirely:
\begin{itemize}
\item In fixed-target experiments luminosities are often high, leading to abundant yields and consequently reduced statistical uncertainties. Moreover, thanks to the boost of the center-of-mass frame of the collision, the rapidity coverage of such experiments can extend up to very large values of rapidity (or Feynman-$x$, $x_{_{\rm F}} \simeq 2\Mt/\sqrts\times \sinh{y} $) using forward spectrometers. However, the highest energies ever reached  are rather modest, $\sqrts=38.7$~GeV and $\sqrts=41.6$~GeV using respectively the 800~GeV and 920~GeV proton beams at the Tevatron and at HERA;
\item At collider energies --~RHIC and LHC, to quote only the facilities accelerating heavy ions~-- unprecedented energies can be reached, respectively $\sqrt{s}=200$~GeV and $\sqrt{s}=5.02$~TeV, making for instance easier the production of  $\Upsilon$ states, marginally measured in fixed-target experiments. In terms of acceptance, quarkonia are detected in a narrow window in $\xf$, centered around $\xf=0$.
\end{itemize}

Let us illustrate this with a few examples, starting with one of the first experiments which measured $\jpsi$ suppression in p--A collisions. The NA3 spectrometer at the CERN SPS collected 1.5~million $\jpsi$ events (!)  in hadron--nucleus collisions~\cite{Badier:1983dg}, allowing for precise measurements close to the kinematic edge of phase-space, $\xf \lesssim 0.75$ (on the contrary, the coverage at RHIC and LHC is respectively  $|\xf|\lesssim0.2$ and $|\xf|\lesssim 0.02$ for $\jpsi$ production). It is remarkable that these data, taken in the early 1980's, prove as competitive as the most recent LHC results when it comes to understand $\jpsi$ suppression in nuclei. More than a decade after NA3, the FNAL E866 experiment reported on high-statistics measurements of $\jpsi$ ($3\times 10^6$ events) and $\psi^\prime$ ($10^5$) production on several nuclear targets in the range $-0.2 \lesssim \xf \lesssim 0.9$~\cite{Leitch:1999ea}. These measurements\footnote{as well as others measurements by, {\it e.g.}, NA50~\cite{Alessandro:2006jt}, NA60~\cite{Arnaldi:2010ky}, and HERA-B~\cite{Abt:2008ya}, yet on a more restricted $\xf$ range.} are nicely supplemented by those carried out in d--Au collisions at RHIC (PHENIX~\cite{Adare:2010fn,Adare:2012qf}, STAR~\cite{Adamczyk:2013poh}) and in p--Pb collisions at LHC (ALICE\cite{Abelev:2013yxa}, LHCb~\cite{Aaij:2013zxa}). At LHC, the relative suppression of $\Upsilon$ excited states (2S, 3S) with respect to 1S states has been performed by CMS~\cite{Chatrchyan:2013nza}, not to mention open heavy-flavor data (D mesons in ALICE~\cite{Abelev:2014hha}, B mesons in CMS~\cite{CMS:2014tfa}, non-prompt $\jpsi$ coming from B decays in ALICE~\cite{Adam:2015rba} and  LHCb~\cite{Aaij:2013zxa}).

Several cold nuclear matter (CNM) effects could in principle affect quarkonium yields in proton--nucleus collisions. Without being comprehensive, let us mention the following ones:
\begin{itemize}
\item Quarkonia may interact inelastically with the surrounding nucleons they may encounter while propagating through the nucleus. 
Such {\it nuclear absorption} may happen when the quarkonium formation time (in the rest frame of the nucleus) is comparable or less than the medium length $L$, $\tau_f \times \cosh \ylab  \lesssim L$, 
where $\tau_f$  is the \emph{proper} formation time
($\tau_f \simeq 0.3$~fm  for both $\jpsi$ and $\Upsilon$),
and $\ylab$ is the quarkonium rapidity in the nucleus rest frame;\footnote{Note that $\ylab$ is directly related to the momentum fraction $x_2$ carried by the target parton, $\cosh{\ylab} = M_\perp / (2m_p x_2)$, where $M_\perp$ is the quarkonium transverse mass and $m_p$ is the proton mass.}
\item Parton distribution functions (PDF) are known to be different in a proton and in a nucleus at all values of $x$. Such nuclear PDF (nPDF) effects could  either suppress or enhance quarkonium yields in p--A collisions (with respect to p--p collisions) depending on the value of the momentum fraction $x_2$. When $x_2$ is small,\footnote{Typically when the time for the hard process to occur is large in the nucleus rest frame, $\thard   \simeq (1/\Mt)\times \cosh \ylab = 1/(2 m_p x_2) \gtrsim L$. Using $L=10$~fm, this would correspond to $x_2 \lesssim 10^{-2}$.} the nucleons in the nucleus act coherently leading to a reduction of the quarkonium yield in a nucleus --~called \emph{shadowing}~\cite{Armesto:2006ph}, or \emph{saturation}~\cite{Gelis:2010nm} to use a more modern language~-- as compared to the incoherent sum over $A$ independent nucleons;
\item Nuclear transverse momentum broadening of the heavy quark pair 
induces coherent gluon radiation, 
arising from the interference between emission amplitudes 
off the initial projectile parton and the final color octet quark pair. 
This coherent medium-induced radiation leads to an average induced 
energy loss proportional to the quarkonium energy~\cite{Arleo:2010rb}. The consequences of coherent energy loss are quarkonium suppression (respectively, enhancement) at large positive (respectively, large negative) values of the rapidity and at all center-of-mass energies of the p--A collision.
\end{itemize}
Obviously it is not because CNM effects \emph{can} play a role that they necessarily do; in particular, the strength of each CNM effect is usually unknown \emph{a priori}. A sound strategy is to investigate each of these effects separately, through a systematic and quantitative comparison to all available data, while keeping the smallest number of  assumptions and free parameters.

Quarkonium suppression reported at forward rapidities cannot be reproduced by either nuclear absorption or nPDF effects, nor by a mixture of both. Although the  comparison to   RHIC and LHC data only may still  give the impression that strong nPDF effects  could explain $\jpsi$ data~\cite{Abelev:2013yxa}\footnote{For examples of nPDF effects on quarkonium production in p--Pb collisions at LHC, see~\cite{Albacete:2013ei,Ferreiro:2013pua}.}, the significant suppression measured by the fixed-target experiments (NA3 and E866) on a wider $\xf$ range  is clearly incompatible with the predictions of these two effects.\footnote{An elegant way to be persuaded is to plot  $\jpsi$ suppression data as a function of $x_2=\Mt/\sqrts \times \exp(-y)$~\cite{Hoyer:1990us}. The suppression from 
either nuclear absorption or nPDF effects is 
expected to be a function of $x_2$ only, independent of $\sqrts$, in violent  disagreement
 with the accumulated data from fixed-target and RHIC experiments (see~\cite{Leitch:1999ea,Leitch:2006ff}).} Without a doubt, the world data indicate that at least another cold nuclear matter effect is at play.

Contrary to nuclear absorption or nPDF effects, 
the sole effect of coherent energy loss is able to reproduce the data on quarkonium suppression, from fixed-target to collider energies~\cite{Arleo:2012hn,Arleo:2012rs,Arleo:2013zua}. Detailed comparisons were published elsewhere, so let us only highlight the phenomenological successes of this approach:
\begin{itemize}
\item The $\xf$ (or, $y$) dependence of $\jpsi$ suppression is well reproduced on a very large domain (up to large values of $\xf \lesssim 0.8$, when data are available) and at all center-of-mass energies, from $\sqrts\simeq 20$~GeV to $\sqrts=5$~TeV;
\item The $\pt$ dependence is well reproduced too, either at a fixed-target experiment (E866) or at colliders (RHIC, LHC), although the $\pt$ dependence seems slightly more abrupt in the model than in collider data. The  centrality dependence measured by PHENIX at RHIC is also nicely described;
\item $\Upsilon$ measurements in p--A collisions are compatible with the expected mass dependence of coherent energy loss, although the present experimental uncertainties are still fairly large;
\item Finally, an original prediction of coherent energy loss is a different 
magnitude of quarkonium suppression 
in p--A and $\pi$--A collisions (in contrast with nuclear absorption effects, which should be independent of the projectile hadron), in agreement with the measurements of NA3.
\end{itemize}
The strength of coherent energy loss depends on a single free parameter, namely the magnitude of the cold nuclear matter transport coefficient, $\hat{q}_0=0.075$~\gevsqfm~at $x=10^{-2}$, obtained from a fit of the precise E866 measurements in p--W collisions.

We find it appealing that the variety of quarkonium measurements in p--A collisions can be described using a \emph{single} CNM effect. Of course, by no means does this  imply that no other CNM effects could play a role too, yet these clearly appear to be subleading when the quarkonium is produced at `large enough' rapidity.
 Both nPDF and coherent energy loss effects could in principle be incorporated consistently in the picture. As a matter of fact, attempts have been made in~\cite{Arleo:2012rs}. However, given the large theoretical uncertainties on nuclear parton distributions\footnote{This is due to the lack of small-$x$ measurements in nuclear collisions. In this respect a high-energy electron-ion collider would be highly beneficial for the community. 
 Let us mention in passing that no coherent energy loss effects are expected in deep-inelastic scattering experiments as the incoming projectile particle is color neutral~\cite{Arleo:2010rb}.} --~especially for gluon densities at small $x$~-- we prefer to focus on the single (but in our opinion, leading) effect of coherent energy loss for which rather precise calculation can be performed.
 
 An exciting possibility to further constrain cold nuclear matter effects 
(on quarkonium production, but not only) 
 would be to smash the LHC proton and lead beams on a collection of fixed nuclear targets~\cite{Brodsky:2012vg}. We believe that this proposal would combine the above-discussed 
advantages of fixed-target \emph{and} collider experiments.

 In this paper, the predictions for quarkonium suppression due to coherent energy loss in p--A collisions at $\sqrts=114.6$~GeV (corresponding to the nominal 7~TeV proton beam energy at the LHC) and Pb--A collisions at $\sqrts=72$~GeV (corresponding to the 2.76~TeV lead beam) are given. Before this, we recall in the next section the main ingredients of our approach.
 
\section{Coherent energy loss model in a nutshell}
\label{sec:model}

\subsection{Formulation}

We briefly detail in this section the basics of the model based on coherent energy loss used to describe $\psi$ 
(denoting $\jpsi$ or $\Upsilon$) 
suppression measured in proton--nucleus collisions.\footnote{The model can also be formulated in heavy-ion (A--B) collision, see~\cite{Arleo:2014oha} for details.}
 The single differential p--A production cross section as a function of the $\psi$ energy reads~\cite{Arleo:2012rs}
\be
\label{eq:xspA0-energy}
\frac{1}{A}\frac{\dd\sigma_{\mathrm{pA}}^{\psi}}{\dd \Ea} \left( \Ea \right)  = \int_0^{\epsamax} \dd \epsa \,{\cal P}(\epsa, \Ea, \ell_{_{\rm A}}^2) \, \frac{\dd\sigma_{\mathrm{pp}}^{\psi}}{\dd \Ea} \left( \Ea+\epsa \right) \, ,
\ee
where $\Ea$ (respectively, $\epsa$) is the energy (respectively, energy loss) of the $Q \bar{Q}$ pair in the rest frame of the nucleus A. The upper limit on the energy loss is $\epsamax=\min\left(\Ea,\Ep-\Ea\right)$, where $\Ep$ is the beam energy in that frame.
${\cal P}$ denotes the energy loss probability distribution, or \emph{quenching weight}.

The quenching weight is related to the medium-induced, coherent radiation spectrum $\dd I/\dd\varepsilon$ given in~\cite{Arleo:2012rs} (and earlier in \cite{Arleo:2010rb}), which is a very good approximation to the exact spectrum computed to all orders in the opacity expansion~\cite{Peigne:2014uha}. It depends on the accumulated transverse momentum transfer $\ell_{_{\rm A}} = \sqrt{\qhat \ellb}$ (assumed to satisfy  $\ell_{_{\rm A}} \ll \mT$) due to soft rescatterings in the nucleus, where $\ellb$ is the medium path-length and $\qhat$ the transport coefficient in cold nuclear matter. More precisely~\cite{Arleo:2012rs},
\be
\label{qhat-model}
\hat{q} \equiv \hat{q}_0 \left[ \frac{10^{-2}}{\min(x_0, x_2)} \right]^{0.3}\ ; \ \ \  x_0 \equiv \frac{1}{2 m_\mathrm{p} \ellb}\ ; \ \ \ x_2 \equiv  \frac{\mT}{\sqrts} \, e^{-y} \, ,
\ee
where $y$ is the quarkonium rapidity in the center-of-mass frame of the proton--nucleon collision.

Using the fact that the quenching weight is a scaling function of the variable $\varepsilon / E$, namely $E\,{\cal P}(\varepsilon,E,\ell^2) = \Phat(\varepsilon/E,\ell^2)$, 
we can rewrite \eq{eq:xspA0-energy} as~\cite{Arleo:2014oha}
\be
\label{eq:xspB}
\frac{1}{A}\frac{\dd\sigma_{\mathrm{pA}}^{\psi}}{\dd y}\left(y, \sqrts \right) = 
\int_0^{\dymax(y)}  \dd\dyb \,
\Phat(e^\dyb-1, \qhat(y)\ellb) \, \, \frac{\dd\sigma_{\mathrm{pp}}^{\psi}}{\dd{y}} \left( y+\dyb, \sqrts \right) \, .
\ee
Here $\dymax(y)= \min\left(\ln{2},y_{\mathrm{max}}-y\right)$, with $y_{\mathrm{max}} = \ln(\sqrts/\mT)$ the maximal $\psi$ rapidity (in the proton--nucleon c.m. frame) allowed by kinematics.
Using~\eq{eq:xspB} 
we can determine 
the nuclear modification factor in  p--A collision,
\be
\label{RpA}
R_{\mathrm{pA}}^{\psi}\left(y, \sqrt{s} \right) = \frac{1}{A} \, {\frac{\dd\sigma_{\mathrm{pA}}^{\psi}}{\dd y} \left(y, \sqrt{s}\right) \biggr/ \frac{\dd\sigma_{\mathrm{pp}}^{\psi}}{\dd y} \left(y, \sqrt{s} \right)} \, .
\ee

As mentioned in the introduction, quarkonium may suffer inelastic interaction with the surrounding nucleons while escaping the nucleus. Although we do not aim to include such an effect in the present predictions, we  nevertheless indicate the critical rapidity $\ycrit$, 
\be
\label{rap-range}
y^{\rm crit}(\sqrt{s}, L)  \equiv \ln\left( \frac{L}{\tau_f} \cdot \frac{2 \mprot}{\sqrts} \right)  \, ,
\ee
below which nuclear absorption might come into play.

\subsection{Ingredients}

The medium length $L$ is obtained from a Glauber model calculation using realistic nuclear densities. The values are given in~\cite{Arleo:2012rs} and reproduced in Table~\ref{tab:L} for the nuclei of interest in the present paper.
In addition, Eq.~\eq{eq:xspB} requires the knowledge of the p--p cross section.
 It is given by a simple parameterization 
$\dd\sigma_{\mathrm{pp}}^{\psi}/ \dd y  \propto \left(1- \frac{2 M_\perp}{\sqrt{s}} \cosh{y} \right)^{n(\sqrts)}$,
where the exponent $n$ is obtained from a fit to p--p measurements. Lacking p--p data at the energies of interest 
($\sqrts=114.6$~GeV in p--A and $\sqrts=72$~GeV in Pb--A collisions), 
an interpolation between the values obtained at FNAL ($\sqrts=38.7$~GeV) and  RHIC ($\sqrts=200$~GeV) energies has been performed. The exponents used in the present paper are given in Table~\ref{tab:expn}. Note that the normalization of the p--p cross section is irrelevant here as it cancels out when computing \eq{RpA}.

\begin{table}[htbp]
{\footnotesize
 \centering
 \begin{tabular}[c]{p{2.3cm}cccc}
   \hline
   \hline
  \multicolumn{4}{c}{}  \\
Nucleus           &  p &{\rm Ca}  & {\rm Cu} &  {\rm Pb} \\[0.3cm]
Atomic mass       &    1& 40   &         63   &   208  \\
$L$ (fm) &1.5& 5.69     &   6.67    & 10.11   \\
&&&& \\
\hline
\hline
\end{tabular}
 \caption{Values of $L$ used in p, Ca, Cu, and Pb targets.}
 \label{tab:L}
 }
\end{table}
%
\begin{table}[htbp]
{\footnotesize
 \centering
 \begin{tabular}[c]{p{2.3cm}cc}
   \hline
   \hline
 & &     \\
Mode    &   Pb--A & p--A \\[0.3cm]
$\sqrt{s}$ (GeV) &72  & 114.6 \\
$n_{_{\jpsi}}$ &  $5.1\pm0.2$ &  $6.0\pm0.3$  \\
$n_{_{\Upsilon}}$ &  $4.1\pm0.3$ &  $5.0\pm0.4$  \\
 & & \\
\hline
\hline
\end{tabular}
 \caption{Values of $n$ used at $\sqrts=72$~GeV and $\sqrts=114.6$~GeV for $\jpsi$ and $\Upsilon$.}
 \label{tab:expn}
 }
\end{table}

The transport coefficient $\qzero$ is the only free parameter of the model. It is determined by fitting the $\jpsi$ suppression measured by E866~\cite{Leitch:1999ea} in p--W over p--Be collisions ($\sqrt{s}=38.7$~GeV), 
see~\cite{Arleo:2012rs}. The obtained value is  $\qzero=0.075^{+0.015}_{-0.005}$~\gevsqfm.
%

\section{Results}
\label{sec:results}

\subsection{p-A mode}
\label{sec:pa}

\begin{figure}[htb]
    \begin{center}
      \includegraphics[width=7.1cm]{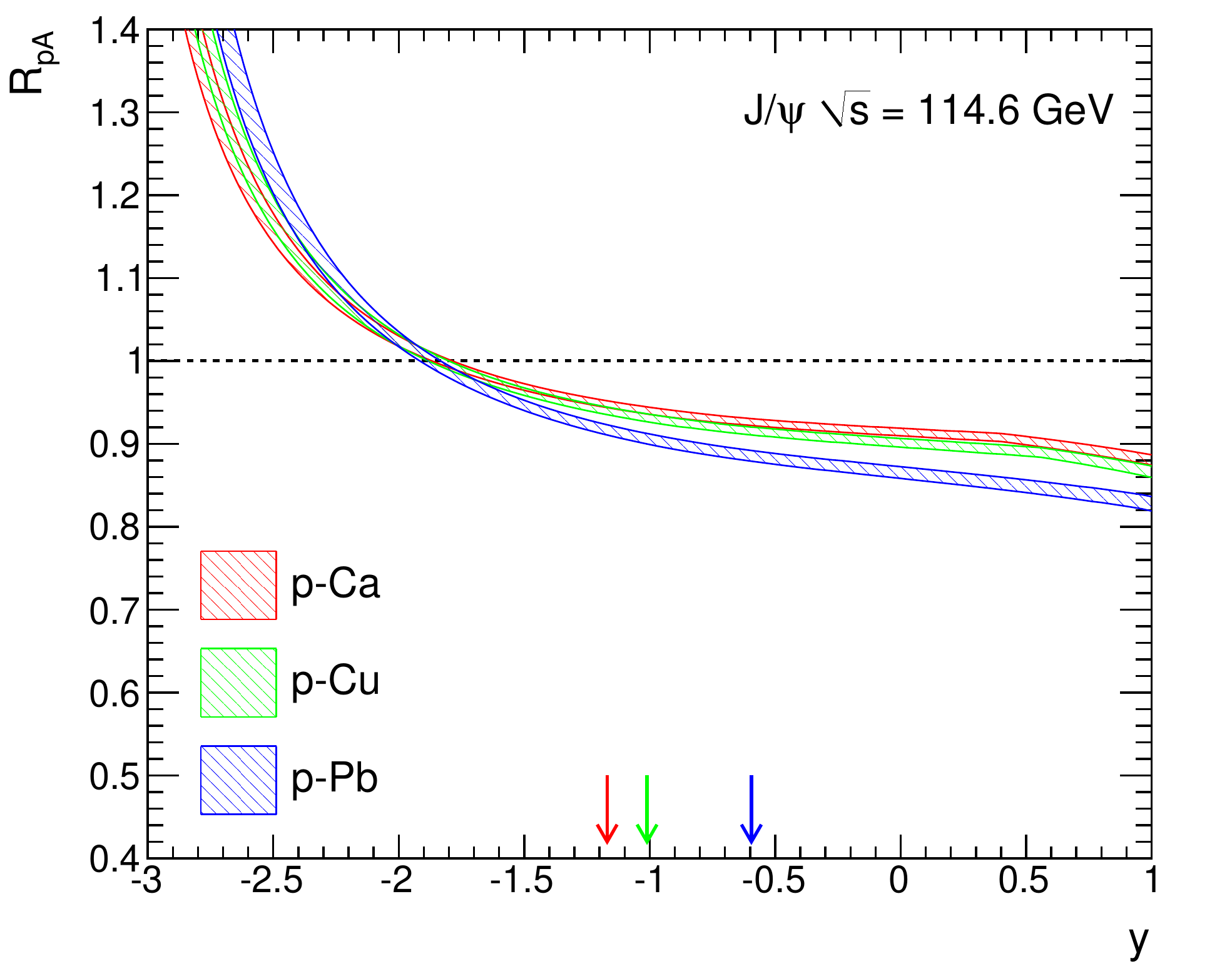}
      \includegraphics[width=7.1cm]{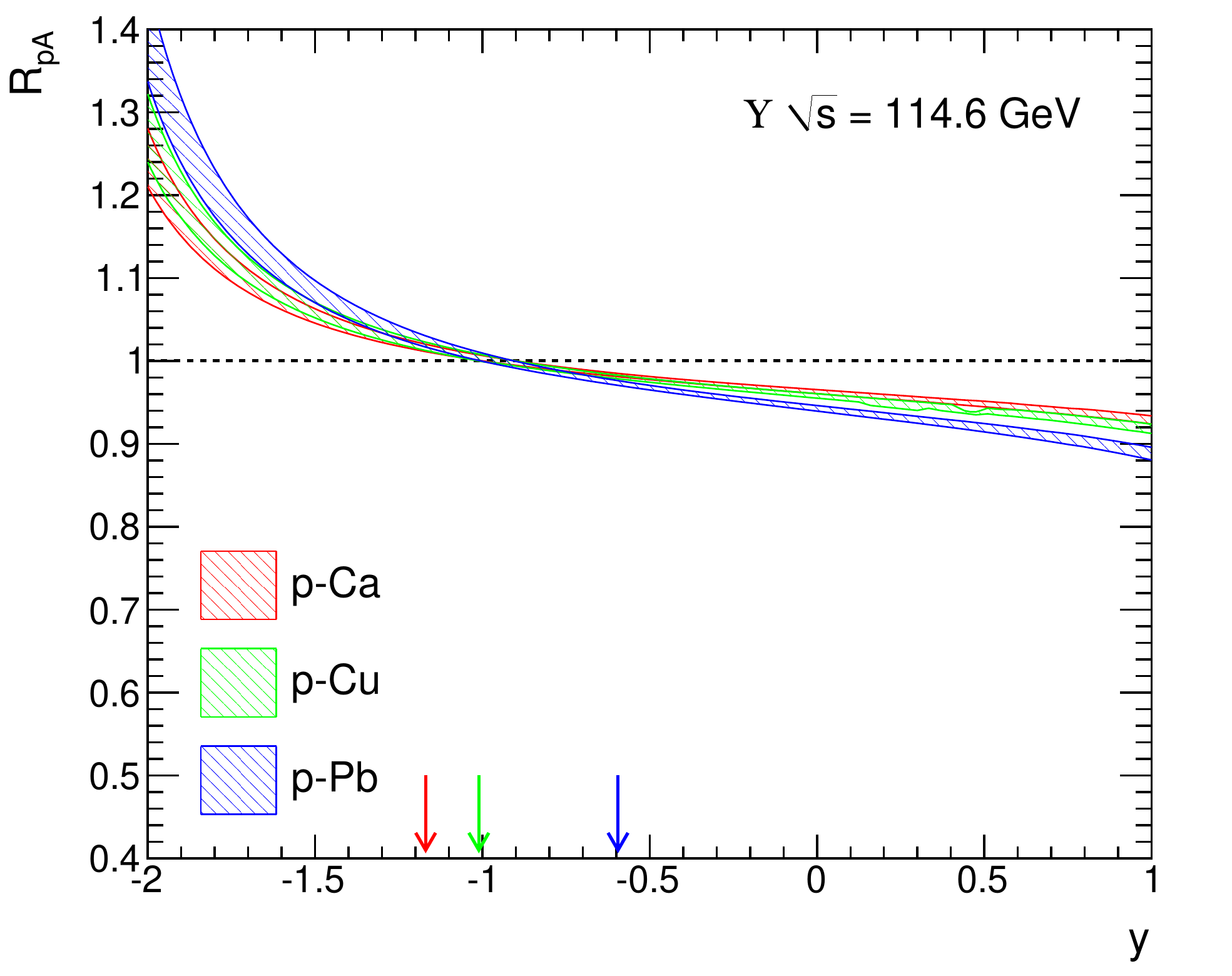}
    \end{center}
    \caption{$\jpsi$ (left) and $\Upsilon$ (right) suppression in p--Ca, p--Cu and p--Pb collisions at $\sqrts=114.6$~GeV.}
    \label{fig:pa}
\end{figure}

The predictions for $\jpsi$ and $\Upsilon$ suppression in p--Ca, p--Cu and p--Pb collisions at $\sqrt{s}=114.6$~GeV are shown in Figure~\ref{fig:pa}. The rapidity range is chosen to match the acceptance of detectors like LHCb. In terms of Feynman-$x$, the range $-3<y<1$ (respectively, $-2<y<1$) correspond to $-0.54 < \xf < 0.06$ (respectively, $-0.60 < \xf < 0.19$) for $\jpsi$ (respectively, $\Upsilon$).

 The $\jpsi$ suppression is rather moderate, less than $20\%$,  and does not vary too strongly with rapidity except at very negative rapidity values, $y < y_0\simeq-2$, where $\jpsi$ enhancement ($\rpa>1$) can be seen. In this rapidity region, however,  nuclear absorption may come into play as can be seen from the vertical arrows indicating the values of $\ycrit$ ($\ycrit \simeq -1$) for each target.

The shape of $\Upsilon$ suppression is similar. The value of the rapidity at which $R_{\rm pA}(y_0)=1$   is  $y_0\simeq-1$, {\it i.e.},  one more unit than in the $\jpsi$ channel. 
This can be understood from the approximate $\xf$ scaling present in the model~\cite{Arleo:2012rs},\footnote{At a given rapidity $y$, the corresponding value of $\xf$ is larger for $\Upsilon$ than for $\jpsi$
 due to the larger transverse mass, $\xf \propto \Mt$.}
 which would predict the difference between these two `crossing points' to be $y_0^{\Upsilon}-y_0^{\jpsi} \sim \ln(M^{\Upsilon}/M^{\jpsi}) \simeq 1.1$.
 Once more, nuclear absorption may affect $\Upsilon$ suppression, although maybe not as much as the $\jpsi$  because of its smaller radius.

\subsection{A-p mode}
\label{sec:ap}

\begin{figure}[htb]
    \begin{center}
      \includegraphics[width=7.1cm]{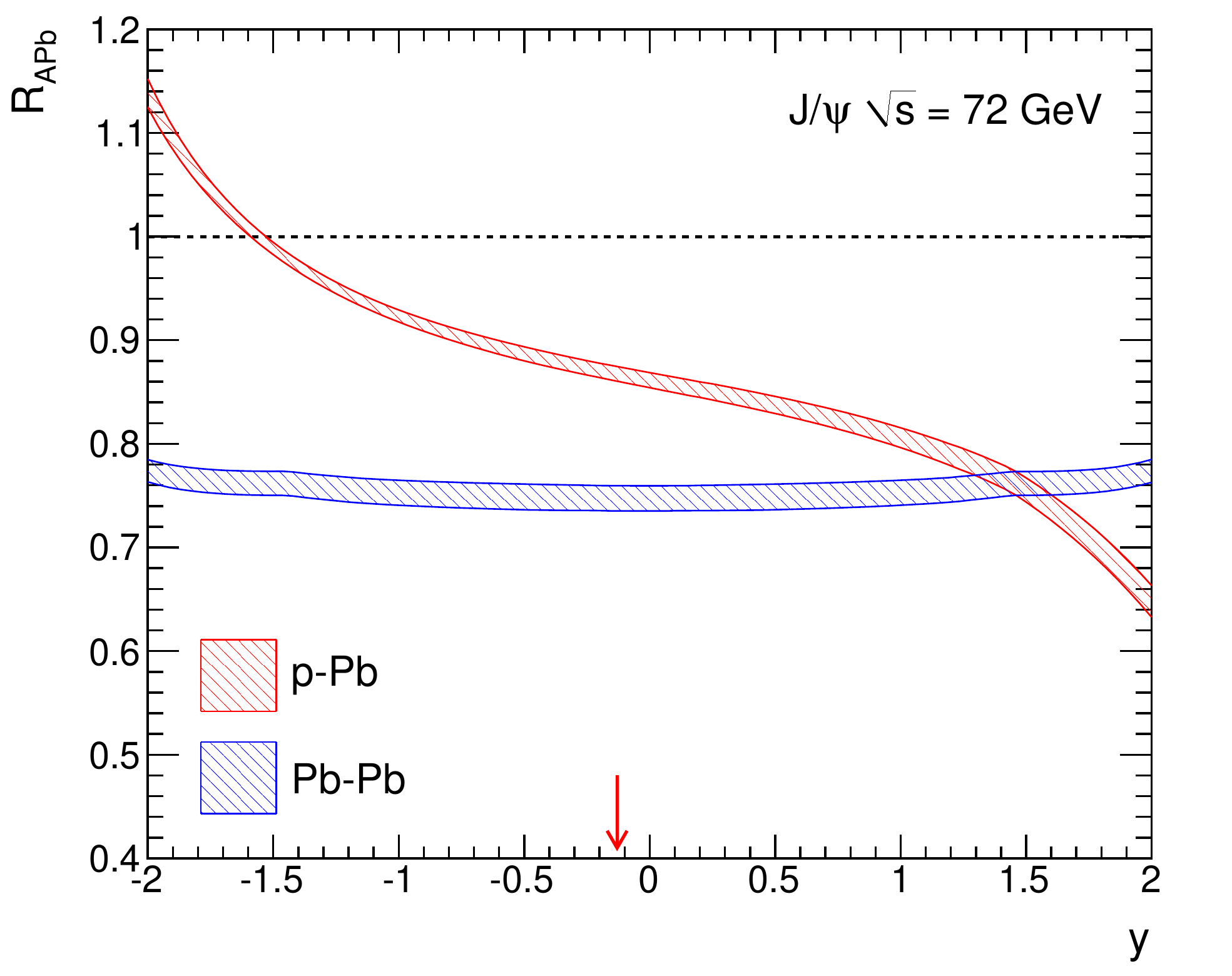}
      \includegraphics[width=7.1cm]{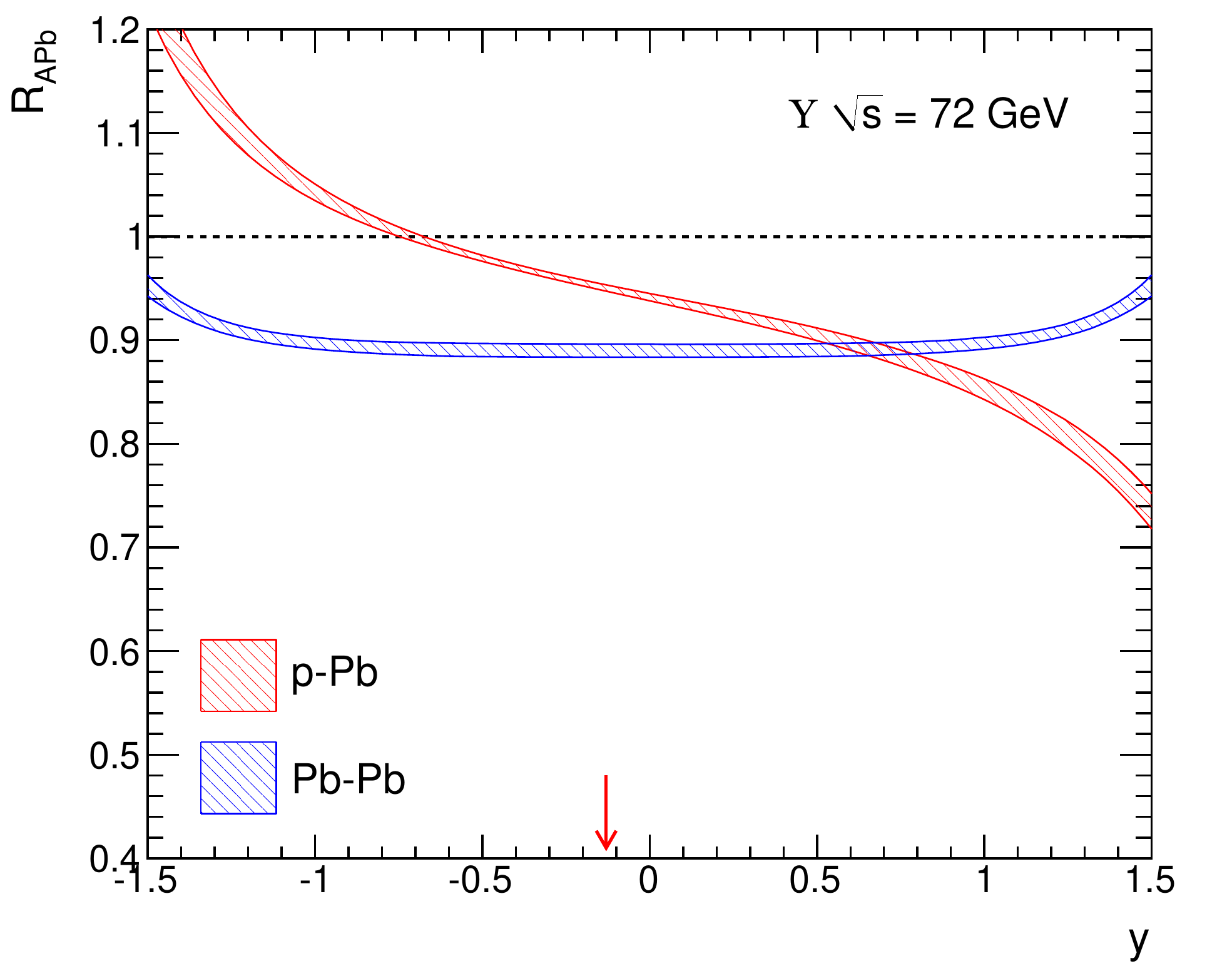}
    \end{center}
    \caption{$\jpsi$ (left) and $\Upsilon$ (right) suppression in p--Pb and Pb--Pb collisions at $\sqrts=72$~GeV.}
    \label{fig:ap}
\end{figure}

Let us move now to calculations corresponding to an incoming 2.76~TeV Pb beam on a proton and a Pb target, 
shown in Figure~\ref{fig:ap}.
 This configuration allows for probing more easily quarkonium suppression in the proton fragmentation, {\it i.e.}, at positive rapidities.\footnote{Although the Pb nucleus collides on a proton, we shall keep the convention
 that positive values of $y$ correspond to the proton fragmentation region.}
  The chosen rapidity range $-2<y<2$ (respectively, $-1.5<y<1.5$) correspond to $-0.31 < \xf < 0.31$ ($-0.56 < \xf < 0.56$) for $\jpsi$ ($\Upsilon$).
The lower center-of-mass energy however shifts $\ycrit$ in p--Pb collisions towards larger values, possibly leading to more pronounced nuclear absorption.

In Pb--Pb collisions the suppression is naturally an even function of $y$. In such collisions, one  expects a hot medium to be formed leading to extra quarkonium suppression. Therefore the results should  rather be
seen as baseline calculations than genuine predictions. Moreover, in A--A collisions the condition for hadronization taking place outside \emph{both} nuclei  reads $\ycrit < y < -\ycrit$. This condition is only met in Pb--Pb collisions at $\sqrts=72$~GeV around mid-rapidity, $|y| \lesssim 0.1$. At larger $|y|$, the quarkonium state shall be fully formed in one of the two nuclei, and thus possibly sensitive to nuclear absorption.

\begin{thebibliography}{10}

\bibitem{Matsui:1986dk}
T.~Matsui and H.~Satz,  Phys. Lett. {\bf B178} (1986) 416.

\bibitem{Badier:1983dg}
{\bf NA3} Collaboration, J.~Badier {\it et~al.},  Z. Phys. {\bf C20} (1983)
  101.

\bibitem{Leitch:1999ea}
{\bf FNAL E866/NuSea} Collaboration, M.~J. Leitch {\it et~al.},  Phys. Rev.
  Lett. {\bf 84} (2000) 3256 [\href{http://arXiv.org/abs/nucl-ex/9909007}{{\tt
  nucl-ex/9909007}}].

\bibitem{Alessandro:2006jt}
{\bf NA50} Collaboration, B.~Alessandro {\it et~al.},  Eur.Phys.J. {\bf C48}
  (2006) 329 [\href{http://arXiv.org/abs/nucl-ex/0612012}{{\tt
  nucl-ex/0612012}}].

\bibitem{Arnaldi:2010ky}
{\bf NA60} Collaboration, R.~Arnaldi {\it et~al.},  Phys. Lett. {\bf B706}
  (2012) 263 [\href{http://arXiv.org/abs/1004.5523}{{\tt 1004.5523}}].

\bibitem{Abt:2008ya}
{\bf HERA-B} Collaboration, I.~Abt {\it et~al.},  Eur. Phys. J. {\bf C60}
  (2009) 525 [\href{http://arXiv.org/abs/0812.0734}{{\tt 0812.0734}}].

\bibitem{Adare:2010fn}
{\bf PHENIX} Collaboration, A.~Adare {\it et~al.},  Phys. Rev. Lett. {\bf 107}
  (2011) 142301 [\href{http://arXiv.org/abs/1010.1246}{{\tt 1010.1246}}].

\bibitem{Adare:2012qf}
{\bf {PHENIX}} Collaboration, A.~Adare, S.~Afanasiev, C.~Aidala, N.~Ajitanand,
  Y.~Akiba {\it et~al.},  \href{http://arXiv.org/abs/1204.0777}{{\tt
  1204.0777}}.

\bibitem{Adamczyk:2013poh}
{\bf STAR} Collaboration, L.~Adamczyk {\it et~al.},  Phys.Lett. {\bf B735}
  (2014) 127 [\href{http://arXiv.org/abs/1312.3675}{{\tt 1312.3675}}].

\bibitem{Abelev:2013yxa}
{\bf ALICE} Collaboration, B.~B. Abelev {\it et~al.},  JHEP {\bf 1402} (2014)
  073 [\href{http://arXiv.org/abs/1308.6726}{{\tt 1308.6726}}].

\bibitem{Aaij:2013zxa}
{\bf LHCb} Collaboration, R.~Aaij {\it et~al.},  JHEP {\bf 1402} (2014) 072
  [\href{http://arXiv.org/abs/1308.6729}{{\tt 1308.6729}}].

\bibitem{Chatrchyan:2013nza}
{\bf CMS} Collaboration, S.~Chatrchyan {\it et~al.},  JHEP {\bf 1404} (2014)
  103 [\href{http://arXiv.org/abs/1312.6300}{{\tt 1312.6300}}].

\bibitem{Abelev:2014hha}
{\bf ALICE} Collaboration, B.~B. Abelev {\it et~al.},  Phys.Rev.Lett. {\bf 113}
  (2014), no.~23 232301 [\href{http://arXiv.org/abs/1405.3452}{{\tt
  1405.3452}}].

\bibitem{CMS:2014tfa}
{\bf CMS} Collaboration,  CMS-PAS-HIN-14-004, \url{http://inspirehep.net/record/1339305/files/HIN-14-004-pas.pdf}

\bibitem{Adam:2015rba}
{\bf ALICE} Collaboration, J.~Adam {\it et~al.},
  \href{http://arXiv.org/abs/1504.07151}{{\tt 1504.07151}}.

\bibitem{Armesto:2006ph}
N.~Armesto,  J. Phys. {\bf G32} (2006) R367
  [\href{http://arXiv.org/abs/hep-ph/0604108}{{\tt hep-ph/0604108}}].

\bibitem{Gelis:2010nm}
F.~Gelis, E.~Iancu, J.~Jalilian-Marian and R.~Venugopalan,
  Ann.Rev.Nucl.Part.Sci. {\bf 60} (2010) 463--489
  [\href{http://arXiv.org/abs/1002.0333}{{\tt 1002.0333}}].

\bibitem{Arleo:2010rb}
F.~Arleo, S.~Peign{\'e} and T.~Sami,  Phys. Rev. {\bf D83} (2011) 114036
  [\href{http://arXiv.org/abs/1006.0818}{{\tt 1006.0818}}].

\bibitem{Albacete:2013ei}
J.~Albacete, N.~Armesto, R.~Baier, G.~Barnafoldi, J.~Barrette {\it et~al.},
  Int.J.Mod.Phys. {\bf E22} (2013) 1330007
  [\href{http://arXiv.org/abs/1301.3395}{{\tt 1301.3395}}].

\bibitem{Hoyer:1990us}
P.~Hoyer, M.~Vanttinen and U.~Sukhatme,  Phys. Lett. {\bf B246} (1990)
  217--220.

\bibitem{Leitch:2006ff}
M.~Leitch,  AIP Conf.Proc. {\bf 892} (2007) 404--409
  [\href{http://arXiv.org/abs/nucl-ex/0610031}{{\tt nucl-ex/0610031}}].

\bibitem{Arleo:2012hn}
F.~Arleo and S.~Peign{\'e},  Phys. Rev. Lett. {\bf 109} (2012) 122301
  [\href{http://arXiv.org/abs/1204.4609}{{\tt 1204.4609}}].

\bibitem{Arleo:2012rs}
F.~Arleo and S.~Peign{\'e},  JHEP {\bf 03} (2013) 122
  [\href{http://arXiv.org/abs/1212.0434}{{\tt 1212.0434}}].

\bibitem{Arleo:2013zua}
F.~Arleo, R.~Kolevatov, S.~Peign{\'e} and M.~Rustamova,  JHEP {\bf 1305} (2013)
  155 [\href{http://arXiv.org/abs/1304.0901}{{\tt 1304.0901}}].

\bibitem{Brodsky:2012vg}
S.~Brodsky, F.~Fleuret, C.~Hadjidakis and J.~Lansberg,  Phys.Rept. {\bf 522}
  (2013) 239--255 [\href{http://arXiv.org/abs/1202.6585}{{\tt 1202.6585}}].

\bibitem{Arleo:2014oha}
F.~Arleo and S.~Peign{\'e},  JHEP {\bf 1410} (2014) 73
  [\href{http://arXiv.org/abs/1407.5054}{{\tt 1407.5054}}].

\bibitem{Peigne:2014uha}
S.~Peign{\'e}, F.~Arleo and R.~Kolevatov,
  \href{http://arXiv.org/abs/1402.1671}{{\tt 1402.1671}}.

\end{thebibliography}

\providecommand{\href}[2]{#2}\begingroup\raggedright\endgroup

\end{document}